\documentstyle[12pt]{article}
\begin{document}\bibliographystyle{plain}
\begin{titlepage}\renewcommand{\thefootnote}{\fnsymbol{footnote}}
\hfill\begin{tabular}{l}HEPHY-PUB
718/99\\UWThPh-1999-45\\hep-ph/9909451\\July 1999\end{tabular}\\[1cm]
\huge\begin{center}{\bf ACCURACY OF APPROXIMATE\\EIGENSTATES}\\
\vspace{1cm}\large{\bf Wolfgang LUCHA\footnote[1]{\normalsize\ {\em
E-mail\/}: wolfgang.lucha@oeaw.ac.at}}\\[.5cm]Institut f\"ur
Hochenergiephysik,\\\"Osterreichische Akademie der
Wissenschaften,\\Nikolsdorfergasse 18, A-1050 Wien, Austria\\[1cm]{\bf Franz
F. SCH\"OBERL\footnote[2]{\normalsize\ {\em E-mail\/}:
franz.schoeberl@univie.ac.at}}\\[.5cm]Institut f\"ur Theoretische
Physik,\\Universit\"at Wien,\\Boltzmanngasse 5, A-1090 Wien,
Austria\vfill{\normalsize\bf Abstract}\end{center}\small Besides perturbation
theory, which requires the knowledge of the exact unperturbed solution,
variational techniques represent the main tool for any investigation of the
eigenvalue problem of some semibounded operator $H$ in quantum theory. For a
reasonable choice of the employed trial subspace of the domain of $H$, the
lowest eigenvalues of $H$ can be located with acceptable precision whereas
the trial-subspace vectors corresponding to these eigenvalues approximate, in
general, the exact eigenstates of $H$ with much less accuracy. Accordingly,
various measures for the accuracy of approximate eigenstates derived by
variational techniques are scrutinized. In particular, the matrix elements of
the commutator of the operator $H$ and (suitably chosen) different operators
with respect to degenerate approximate eigenstates of $H$ obtained~by~the
variational methods are proposed as new criteria for the accuracy of
variational~eigenstates. These considerations are applied to that Hamiltonian
the eigenvalue problem of which defines the spinless Salpeter equation. This
bound-state wave equation may be regarded as the most straightforward
relativistic generalization of the usual nonrelativistic Schr\"odinger
formalism, and is frequently used to describe, e.g., spin-averaged mass
spectra of bound states of quarks.

\vspace{3ex}

\noindent{\em PACS\/}: 03.65.Ge, 03.65.Pm, 11.10.St, 12.39.Ki
\renewcommand{\thefootnote}{\arabic{footnote}}\end{titlepage}

\normalsize

\section{Motivation}A central element in quantum theory is the solution of
eigenvalue problems. However, usually there is no suitable exact solution to
which perturbation theory can~be~applied.

A very efficient way to locate the discrete spectrum of some self-adjoint
operator~$H$ bounded from below is provided by the famous Rayleigh--Ritz
variational technique~\cite{MMP}: {\sl If the eigenvalues $E_k$,
$k=0,1,\dots,$ of $H$ are ordered according to $E_0\le E_1\le E_2\le\dots,$
the first $d$ of them are bounded from above by the $d$ eigenvalues $\widehat
E_k$, $k=0,1,\dots,d-1,$ (ordered by $\widehat E_0\le\widehat
E_1\le\dots\le\widehat E_{d-1}$) of that operator which is obtained by
restricting~$H$ to some $d$-dimensional subspace of the domain of $H$, i.e.,
$E_k\le\widehat E_k$, $k=0,1,\dots,d-1.$} Unless the eigenstate
$|\chi_k\rangle$ corresponding to some eigenvalue $E_k$ is already an
element~of this trial space---in which case the upper bound $\widehat E_k$
becomes identical to the eigenvalue $E_k$---, enlarging the dimension $d$ of
the trial space will, in general, improve the obtained upper bound; at least,
it can't make things worse. However, it is not straightforward~to quantify
how close approximate eigenstates resulting from the variational method and
exact eigenstates are.\footnote{\ Recent investigations of the production and
decay of heavy quarkonium systems have stimulated renewed interest in the
exact value of the wave function at the origin of the two-quark bound
state~\cite{WFO}.} Thus, we embark upon a systematic study of the accuracy
of~the variationally determined eigenstates of $H$ and suitable measures to
judge their quality.

\section{Measures of Quality for Variational Trial
States}\label{Sec:MQ}Consider some self-adjoint operator $H$, $H^\dagger=H$,
assumed to be bounded from below. Suppressing, for the moment, the index
$k=0,1,2,\dots$ discriminating between different solutions, let the
eigenvalue equation for $H$,
\begin{equation}H|\chi\rangle=E|\chi\rangle\ ,\label{eq:GEVE}\end{equation}be
solved by some (generic) eigenvector $|\chi\rangle$ corresponding to some
(real) eigenvalue~$E$, to be extracted according to
$$E\equiv\frac{\langle\chi|H|\chi\rangle}{\langle\chi|\chi\rangle}\ .$$The
Rayleigh--Ritz variational technique yields an upper bound $\widehat E$ on
this eigenvalue~$E$ as well as, by diagonalization of the relevant
characteristic equation, the corresponding vector $|\varphi\rangle$ in the
$d$-dimensional trial space. There exist several (potentially meaningful)
measures~of the quality of this trial state $|\varphi\rangle$ which
immediately come to one's mind:
\begin{enumerate}\item The trial state $|\varphi\rangle$ is supposed to
represent---to a certain degree of accuracy---the approximate solution of the
eigenvalue problem defined in Eq.~(\ref{eq:GEVE}). Consequently, a first
indicator for the resemblance of $|\varphi\rangle$ with the exact eigenstate
$|\chi\rangle$ would~be the distance between the expectation value of the
operator $H$ with respect to~the trial state $|\varphi\rangle$, i.e., between
the obtained upper bound$$\widehat E\equiv
\frac{\langle\varphi|H|\varphi\rangle}{\langle\varphi|\varphi\rangle}\ ,$$
and the exact eigenvalue $E$. However, the precise location of the exact
eigenvalue $E$ is usually not known. Lower bounds on eigenvalues of
self-adjoint operators are much harder to find than upper bounds. The
practical use of Temple's~inequality \cite{MMP} is, unfortunately, limited.
The ``local-energy'' theorem \cite{LET} applies, in general, only to the
ground state.\newpage
\item The natural measure for the resemblance of the Hilbert-space vectors
$|\varphi\rangle$ and~$|\chi\rangle$ under consideration is the overlap
\begin{equation}S\equiv\frac{\langle\varphi|\chi\rangle}
{\sqrt{\langle\varphi|\varphi\rangle\,\langle\chi|\chi\rangle}}\label{eq:OTE}
\end{equation}of the trial state $|\varphi\rangle$ with the eigenstate
$|\chi\rangle$. If the generic eigenstate $|\chi\rangle$ refers~to the ground
state $|\chi_0\rangle$ of the operator $H$, the deviation of this overlap
from unity is, according to the (rather efficient) Eckart criterion
\cite{Eckart30}, bounded from above~by\begin{equation}
1-\frac{|\langle\varphi|\chi_0\rangle|^2}{\langle\varphi|\varphi\rangle\,
\langle\chi_0|\chi_0\rangle}\le\frac{1}{E_1-E_0}
\left(\frac{\langle\varphi|H|\varphi\rangle}{\langle\varphi|\varphi\rangle}
-E_0\right)\equiv\frac{\widehat E_0-E_0}{E_1-E_0}\
.\label{eq:EC}\end{equation}A comprehensive discussion of both lower and
upper bounds on the overlap $S$ as criteria for the accuracy of approximate
wave functions may be found in~Ref.~\cite{Weinhold70}. As is evident even
from the simple Eckart criterion (\ref{eq:EC}), all these bounds require, in
general, some additional knowledge, such as the location of the
eigenvalues~$E_k$, $k=1,2,\dots,$ of the operator $H$ or matrix elements of
higher powers $H^n$, $n\ge2$,~of $H$, for instance, certain moments
$$\frac{\langle\varphi|H^n|\varphi\rangle}{\langle\varphi|\varphi\rangle}\
,\quad n\ge2\ ,$$of $H$, or, at least, appropriate bounds on these quantities
\cite{Weinhold70}. Rather frequently, however, the required information is
either not available at all or hard to obtain.
\item Consider the commutator $[G,H]$ of the operator $H$ under consideration
with~any other operator $G$, where the domain of $G$ is assumed to contain
the domain~of~$H$. Then the expectation value of this commutator with respect
to a given eigenstate $|\chi\rangle$ of $H$ (or, more generally, the matrix
element of this commutator with respect to an arbitrary pair of {\em
degenerate\/} eigenstates $|\chi_i\rangle$ and $|\chi_j\rangle$ of $H$, i.e.,
eigenstates which satisfy $E_i=E_j$) trivially vanishes:
\begin{equation}\langle\chi|[G,H]|\chi\rangle=0\ .\label{eq:VC}\end{equation}
(Under the name ``hypervirial theorems'' \cite{HVT}, this relation has been
generalized, for Hamiltonians $H$ which involve nonrelativistic kinematics,
to nonnormalizable vectors $|\chi\rangle$, that is, to vectors which are not
elements of some Hilbert space \cite{NNV}.) Hence, choosing different
operators $G$ generates a whole class of operators $[G,H]$ each of which may
serve to test the quality of a given trial state $|\varphi\rangle$ by
evaluating how close the expectation value
$\langle\varphi|[G,H]|\varphi\rangle$ of this operator with respect to
$|\varphi\rangle$ comes to
zero:$$\langle\varphi|[G,H]|\varphi\rangle\stackrel{?}{\simeq}0\ .$$This
expectation value vanishes, of course, also if, by accident, the state
$|\varphi\rangle$~is~an eigenstate of $G$. However, for a given operator $G$,
after having determined $|\varphi\rangle$,~it is straightforward to check for
this circumstance, for instance, by inspecting the variance $\Delta_G$ of the
operator $G$ with respect to the state $|\varphi\rangle$, defined, as
usual,~by$$\Delta_G\equiv
\frac{\langle\varphi|G^2|\varphi\rangle}{\langle\varphi|\varphi\rangle}-
\left(\frac{\langle\varphi|G|\varphi\rangle}{\langle\varphi|\varphi\rangle}
\right)^2\ ,$$which clearly vanishes if the state $|\varphi\rangle$ under
consideration is an eigenstate of $G$. Moreover, it goes without saying that
an expectation value $\langle\varphi|[G,H]|\varphi\rangle$ vanishes also if
the state $|\varphi\rangle$ is an eigenstate of the commutator $[G,H]$ with
eigenvalue~0, or even if the state defined by $[G,H]|\varphi\rangle$ proves
to be orthogonal to the state~$|\varphi\rangle$.

For any self-adjoint operator $G$, i.e., $G^\dagger=G$, this commutator is
anti-Hermitean, which obviously suggests to define a self-adjoint operator
$C=C^\dagger$ (on the domain of the operator $H$) by$$[G,H]=:{\rm i}\,C\
.$$If, for example, $G$ is chosen to be the (symmetrized and self-adjoint)
generator~of dilations,\begin{equation}G\equiv\frac{1}{2}\,({\bf x}\cdot{\bf
p}+{\bf p}\cdot{\bf x})\ ,\label{eq:DG}\end{equation}the relation
(\ref{eq:VC}) is precisely the ``master virial theorem'' introduced in
Ref.~\cite{Lucha90:RVTs}~for a systematic study of (relativistic) virial
theorems \cite{Lucha89:RVT}. In this case, for operators $H$ of the form of
some typical Hamiltonian consisting of a momentum-dependent kinetic-energy
operator, $T({\bf p})$, and a coordinate-dependent interaction-potential
operator, $V({\bf x})$, that is,$$H=T({\bf p})+V({\bf x})\ ,$$the operator
$C$ becomes the ``virial operator''\begin{equation}C={\bf
p}\cdot\frac{\partial}{\partial{\bf p}}T({\bf p})-{\bf
x}\cdot\frac{\partial}{\partial{\bf x}}V({\bf x})\
.\label{eq:VO}\end{equation}The point spectrum (i.e., the set of all
eigenvalues) of the dilation generator~(\ref{eq:DG}) is empty; in other
words, the dilation generator has no Hilbert-space eigenvectors.
\end{enumerate}In any case, however, one should always keep in mind one
(trivial) fact \cite{Messiah}: Depending on one's particular choice of the
employed trial subspace, it may happen that one trial state
$|\varphi^{(1)}\rangle$ yields a better approximation to the exact eigenstate
$|\chi\rangle$, as judged by~their
overlap$$\frac{\langle\varphi^{(1)}|\chi\rangle}
{\sqrt{\langle\varphi^{(1)}|\varphi^{(1)}\rangle\,\langle\chi|\chi\rangle}}\
,$$but a worse upper bound $\widehat E^{(1)}$ on the corresponding eigenvalue
$E$ while some other~trial state $|\varphi^{(2)}\rangle$ yields the better
upper bound $\widehat E^{(2)}<\widehat E^{(1)}$ on the eigenvalue $E$ but a
worse approximation to the exact eigenstate $|\chi\rangle$, which fact would
be betrayed by the overlap$$\frac{\langle\varphi^{(2)}|\chi\rangle}
{\sqrt{\langle\varphi^{(2)}|\varphi^{(2)}\rangle\,\langle\chi|\chi\rangle}}<
\frac{\langle\varphi^{(1)}|\chi\rangle}
{\sqrt{\langle\varphi^{(1)}|\varphi^{(1)}\rangle\,\langle\chi|\chi\rangle}}\
.$$

\section{Prototype of Relativistic Wave Equations: The Spinless Salpeter
Equation}\label{Sec:SSE}Let us apply the above general considerations to the
prototype of all (semi-) relativistic bound-state equations, the ``spinless
Salpeter equation,'' defined by a---by assumption, self-adjoint---Hamiltonian
$H$ (in one-particle form, which encompasses the equal-mass two-particle
case, too \cite{Lucha94varbound,Lucha96:AUB,Lucha98O,Lucha98D}):
\begin{equation}H=T+V\ ,\label{Eq:SRH}\end{equation}where $T$ is the
``square-root'' operator of the relativistic kinetic energy of some particle
of mass $m$ and momentum ${\bf p}$,\begin{equation}T=T({\bf
p})\equiv\sqrt{{\bf p}^2+m^2}\ ,\label{eq:RKE}\end{equation}and $V=V({\bf
x})$ is an arbitrary, coordinate-dependent, static interaction potential. The
spinless Salpeter equation is then just the eigenvalue equation for this
Hamiltonian~$H$,$$H|\chi_k\rangle=E_k|\chi_k\rangle\ ,\quad k=0,1,2,\dots\
,$$for the complete set of Hilbert-space eigenvectors $|\chi_k\rangle$ of $H$
corresponding to its~energy eigenvalues
$$E_k\equiv\frac{\langle\chi_k|H|\chi_k\rangle}{\langle\chi_k|\chi_k\rangle}\
.$$Analytic upper bounds on these eigenvalues have been given
\cite{Lucha94varbound,Lucha96:AUB,Lucha98O,Lucha98D,Lucha96:CCC,Lucha97:L,
Lucha98R}. For the Coulomb potential, the local-energy theorem has been
successfully applied~\cite{Raynal94}.

For the sake of comparison, we focus our interest to central potentials
$V({\bf x})=V(r)$, $r\equiv|{\bf x}|$. Furthermore, in order to facilitate
the numerical treatment of the problem,~we consider the harmonic-oscillator
potential\begin{equation}V(r)=a\,r^2\ ,\quad a>0\
.\label{eq:HOP}\end{equation}The reason for this particular choice is the
following: In momentum space, the operator $r^2$ is represented by the
Laplacian with respect to the momentum~${\bf p}$, $r^2\rightarrow-\Delta_{\bf
p}$, while the kinetic energy $T$, nonlocal in configuration space, is
represented by a multiplication operator. Consequently, exactly for a
harmonic-oscillator potential the semirelativistic Hamiltonian $H$ in its
momentum-space representation is equivalent to a nonrelativistic Hamiltonian
with some (effective) interaction potential reminiscent of the square
root:\begin{equation}H=-a\,\Delta_{\bf p}+\sqrt{{\bf p}^2+m^2}\
.\label{eq:NRH}\end{equation}The solutions of the corresponding eigenvalue
equation may then be found with one of the numerous procedures designed for
the treatment of the nonrelativistic Schr\"odinger equation.

For the harmonic-oscillator potential, it is comparatively easy to get a
first idea~of the approximate location of the energy levels $E_k$ by entirely
analytical considerations:\begin{itemize}\item On the one hand, since$$T({\bf
p})\le m+\frac{{\bf p}^2}{2\,m}\ ,$$every energy eigenvalue $E_k$ of $H$ is
bounded from above by the energy eigenvalue $E_{k,{\rm NR}}$ of the
nonrelativistic counterpart$$H_{\rm NR}=m+\frac{{\bf p}^2}{2\,m}+V$$ of
$H$:$$E_k\le E_{k,{\rm NR}}\ .$$For the harmonic-oscillator potential
(\ref{eq:HOP}), these nonrelativistic energy levels
read\begin{equation}E_{\rm NR}=m+\sqrt{\frac{2\,a}{m}}\,\left(2\,n_{\rm
r}+\ell+\frac{3}{2}\right),\label{Eq:NREL}\end{equation}expressed in terms of
radial quantum number $n_{\rm r}=0,1,2,\dots$ and orbital angular momentum
$\ell=0,1,2,\dots.$ An (easily) improved bound may be found in
App.~\ref{App:RHOP}.\item On the other hand, since$$T({\bf p})\ge|{\bf p}|\
,$$every energy eigenvalue $E_k$ of $H$ is bounded from below by the energy
eigenvalue $E_k(m=0)$ of the Hamiltonian$$H(m=0)=|{\bf p}|+V$$corresponding
to vanishing particle mass $m$:$$E_k\ge E_k(m=0)\ .$$For the
harmonic-oscillator potential (\ref{eq:HOP}), by remembering the
momentum-space representation (\ref{eq:NRH}) of $H$ all bound states of
vanishing orbital angular momentum $\ell$ (``S waves'') may be obtained as
solutions of a differential equation of the form$$\frac{{\rm d}^2}{{\rm
d}z^2}f(z)-z\,f(z)=0\ ;$$the resulting energy levels
read$$E(m=0)=-a^{1/3}\,z_0\ ,$$where $z_0$ denote the zeros of the Airy
function ${\rm Ai}(z)$, which solves this differential equation:
$z_0=-2.33810\dots,\,-4.08794\dots,\,-5.52055\dots,\,\dots$
\cite{Abramowitz}. Energy~levels of bound states of nonvanishing orbital
angular momentum $\ell$ (``P, D, \dots waves'') must be determined
numerically anyway. Also this bound is improved in
App.~\ref{App:RHOP}.\end{itemize}

For homogeneous functions, the radial derivatives entering in the virial
operator~(\ref{eq:VO}) simply probe the respective degree of homogeneity of
these functions. In particular,~for the harmonic-oscillator potential
(\ref{eq:HOP}) one finds$${\bf x}\cdot\frac{\partial}{\partial{\bf x}}V({\bf
x})=r\,\frac{\partial}{\partial r}V(r)=2\,a\,r^2=2\,V(r)\ .$$With the
expression (\ref{eq:VO}) for $C$, introducing the (only momentum-dependent)
operator$$F\equiv T({\bf p})+\frac{1}{2}\,{\bf
p}\cdot\frac{\partial}{\partial{\bf p}}T({\bf p})\ ,$$the Hamiltonian
(\ref{Eq:SRH}) with harmonic-oscillator potential thus may be cast into the
form\begin{eqnarray*}H&=&T({\bf p})+V({\bf x})\\[1ex]&=&T({\bf
p})+\frac{1}{2}\,{\bf x}\cdot\frac{\partial}{\partial{\bf x}}V({\bf
x})\\[1ex]&=&T({\bf p})+\frac{1}{2}\,{\bf p}\cdot\frac{\partial}{\partial{\bf
p}}T({\bf p})-\frac{C}{2}\,\\[1ex]&\equiv&F-\frac{C}{2}\ .\end{eqnarray*}This
observation allows to express any matrix element of the operator $C$ (in
particular, its expectation value $\langle\varphi|C|\varphi\rangle$ with
respect to the trial state $|\varphi\rangle$ under investigation)~as the
corresponding matrix element of the difference of Hamiltonian $H$ and
operator $F$:
$$\langle\varphi|C|\varphi\rangle=2\,\langle\varphi|F-H|\varphi\rangle\equiv
2\left(\langle\varphi|F|\varphi\rangle-\widehat
E\,\langle\varphi|\varphi\rangle\right).$$For the relativistic expression
(\ref{eq:RKE}) of the kinetic energy $T$, the above operator $F$
reads$$F=\frac{3\,{\bf p}^2+2\,m^2}{2\,\displaystyle\sqrt{{\bf p}^2+m^2}}\
.$$

\section{Definition of the ``Laguerre'' Trial Space}As far as the achieved
accuracy of the solutions obtained is concerned, the most crucial step in all
variational games of the Rayleigh--Ritz kind is, for a given operator $H$
under consideration, a reasonable definition of the adopted trial subspace of
the~domain~of~$H$.

For spherically symmetric potentials $V(r)$, a very popular choice for the
basis states which span the trial space required for the application of the
variational technique are ``Laguerre'' trial states, given in
configuration-space representation by \cite{LTS,Lucha97:L,Lucha98O,Lucha98D}
\begin{equation}\psi_{k,\ell m}({\bf
x})=\sqrt{\frac{(2\,\mu)^{2\,\ell+2\,\beta+1}\,k!}
{\Gamma(2\,\ell+2\,\beta+k+1)}}\,r^{\ell+\beta-1}\exp(-\mu\,r)\,
L_k^{(2\,\ell+2\,\beta)}(2\,\mu\,r)\,{\cal Y}_{\ell m}(\Omega_{\bf x})\
,\label{eq:LTF}\end{equation} where $L_k^{(\gamma)}(x)$ are the generalized
Laguerre polynomials (for parameter $\gamma$) \cite{Abramowitz},~defined by
the power
series$$L_k^{(\gamma)}(x)=\sum_{t=0}^k\,(-1)^t\left(\begin{array}{c}
k+\gamma\\k-t\end{array}\right)\frac{x^t}{t!}$$and normalized, with the
weight function $x^\gamma\exp(-x)$, according to$$\int\limits_0^\infty{\rm
d}x\,x^\gamma\exp(-x)\,L_k^{(\gamma)}(x)\,
L_{k'}^{(\gamma)}(x)=\frac{\Gamma(\gamma+k+1)}{k!}\,\delta_{kk'}\ ,$$ and
${\cal Y}_{\ell m}(\Omega)$ are the spherical harmonics for angular momentum
$\ell$ and its projection~$m$, depending on~the solid angle $\Omega$ and
orthonormalized according to $$\int{\rm d}\Omega\,{\cal Y}^\ast_{\ell
m}(\Omega)\,{\cal Y}_{\ell'm'}(\Omega)=\delta_{\ell\ell'}\,\delta_{mm'}\ .$$
The trial functions (\ref{eq:LTF}) involve two variational parameters, $\mu$
(with dimension of~mass) and $\beta$ (dimensionless), which, by the
requirement of normalizability of these functions, are subject to the
constraints $\mu>0$ and $2\,\beta>-1$.

One of the advantages of the trial function (\ref{eq:LTF}) is the easy
availability of an~analytic expression for the corresponding momentum-space
representation of these trial states, obtained by Fourier
transformation:\begin{eqnarray*}\widetilde\psi_{k,\ell m}({\bf p})
&=&\sqrt{\frac{(2\,\mu)^{2\,\ell+2\,\beta+1}\,k!}
{\Gamma(2\,\ell+2\,\beta+k+1)}}\,\frac{(-{\rm i})^\ell\,|{\bf
p}|^\ell}{2^{\ell+1/2}\,\Gamma\left(\ell+\frac{3}{2}\right)}\,\sum_{t=0}^k\,
\frac{(-1)^t}{t!}\left(\begin{array}{c}k+2\,\ell+2\,\beta\\
k-t\end{array}\right)\\[1ex]
&\times&\frac{\Gamma(2\,\ell+\beta+t+2)\,(2\,\mu)^t}{({\bf
p}^2+\mu^2)^{(2\,\ell+\beta+t+2)/2}}\,
F\left(\frac{2\,\ell+\beta+t+2}{2},-\frac{\beta+t}{2};\ell+\frac{3}{2};
\frac{{\bf p}^2}{{\bf p}^2+\mu^2}\right)\\[1ex]
&\times&{\cal Y}_{\ell m}(\Omega_{\bf p})\ ,\end{eqnarray*}with the
hypergeometric series $F$, defined by
$$F(u,v;w;z)=\frac{\Gamma(w)}{\Gamma(u)\,\Gamma(v)}\,\sum_{n=0}^\infty\,
\frac{\Gamma(u+n)\,\Gamma(v+n)}{\Gamma(w+n)}\,\frac{z^n}{n!}\ .$$

For the present investigation, we too employ the ``Laguerre'' trial states
defined by Eq.~(\ref{eq:LTF}), with, for both definiteness and ease of
calculation, the variational parameters $\mu$ and $\beta$ kept fixed to the
values $\mu=m$ and $\beta=1$.\footnote{\ Within some specific nonrelativistic
quark potential model, the significance of the wave function at the origin
obtained variationally from superpositions of Gaussian trial functions has
been discussed numerically \cite{NRQPM}.}

Of course, the trivial fact that the (Hilbert-space) states defined by the
``Laguerre'' trial functions (\ref{eq:LTF}) cannot be eigenvectors of the
dilation generator (\ref{eq:DG}) is reflected~by the nonvanishing numerical
value of the variance $\Delta_G$ of the operator $G$ in these states: For
instance, for the lowest-dimensional possibility, $d=1$, realized by the
trial function$$\psi_{0,00}({\bf x})=\sqrt{\frac{\mu^3}{\pi}}\exp(-\mu\,r)\
,$$this variance is exactly $\Delta_G=3/4$. For trial-space dimension $d=25$,
this quantity~takes in the state corresponding to the lowest eigenvalue
$\widehat E$ the numerical value $\Delta_G=1.3779$.

Up to a dimension $d=4$ of the trial space, the matrix elements of the
Hamiltonian $H$ with respect to the chosen trial states are, at least in
principle, accessible by entirely algebraic manipulations. In the simplest
conceivable case, realized by the choice $d=1$, one finds with the help of
the explicit analytic expressions for the matrix elements~of~$H$ derived in
Ref.~\cite{Lucha97:L}, for instance, for the ground-state energy $E_0$ the
naive upper bound$$\widehat E_0=\frac{64\,m}{15\,\pi}+\frac{3\,a}{m^2}\ .$$

\section{Rates of Convergence of the Quality Measures}Now, let us observe our
variational eigenstates, $|\varphi\rangle$, approaching the exact
eigenstates, $|\chi\rangle$, for increasing dimension $d$ of the employed
trial space, by comparing the~behaviour of the various measures for the
accuracy of approximate eigenstates discussed in Sec.~\ref{Sec:MQ}.

Without doubt, the only genuine ``point of reference'' of any variational
solution~to an eigenvalue problem is the corresponding exact solution. The
exact solution sought~is computed here with the help of the numerical
integration procedure developed for the solution of the nonrelativistic
Schr\"odinger equation in Ref.~\cite{Falkenst85}. (Since the problematic term
in our Hamiltonian is the kinetic-energy operator (\ref{eq:RKE}), the
solution is constructed in momentum space,\footnote{\ The outcome of any
numerical integration procedure is a solution given only at discrete
points~in the respective representation space. The reconstruction of the
entire function, with the help of some interpolation procedure, in form of an
interpolating function introduces, clearly, an additional error. We have, of
course, checked that this error is negligible.} which, in addition, allows to
take advantage of the operator $F$.)

Table~\ref{Tab:QVS} confronts, for the ground state and the lowest radial and
orbital excitations, the approximate solutions as calculated with the help of
the Rayleigh--Ritz variational technique for ``Laguerre'' trial subspaces of
the domain of $H$ of increasing dimension~$d$ with the exact solutions of the
eigenvalue problem for the semirelativistic Hamiltonian (\ref{Eq:SRH}) with a
central interaction potential of the harmonic-oscillator form (\ref{eq:HOP}).
First of~all, as discussed in Sec.~\ref{Sec:SSE}, the exact position of any
eigenvalue $E$ of our Hamiltonian $H$ is confined to a range defined by the
nonrelativistic upper bound $E_{\rm NR}$ and the~zero-mass lower bound
$E(m=0)$ on this energy eigenvalue $E$. There are several quantities~which
may participate in a competition for ``the best or most reasonable measure
of quality:''\begin{enumerate}\item The relative error $\varepsilon$ of every
(Rayleigh--Ritz) upper bound $\widehat E$ on the exact energy eigenvalue
$E$,\begin{equation}\varepsilon\equiv\frac{\widehat E-E}{E}\
,\label{eq:REE}\end{equation}is, by definition, always nonnegative, i.e.,
$$\varepsilon\ge0\ .$$

\small\begin{table}[ht]\caption[]{\small Characterization of the quality of
the variational solution of the eigenvalue problem of the semirelativistic
Hamiltonian $H=\sqrt{{\bf p}^2+m^2}+V(r)$ with harmonic-oscillator potential
$V(r)=a\,r^2$, for states of radial quantum number $n_{\rm r}=0,1,2$ and
orbital angular momentum $\ell=0,1,2$ (called 1S, 2S, 3S, 1P, and 1D in usual
spectroscopic notation), obtained with the help of ``Laguerre'' trial states
spanning trial spaces of increasing dimension $d=1,2,3,10,25$, by: the
nonrelativistic upper bound $E_{\rm NR}$ and zero-mass lower bound $E(m=0)$
on the energy, the (numerically computed) ``exact'' energy $E$, the
variational upper bound $\widehat E$ on this energy, the relative error
$\varepsilon$ of the upper bound, the deviation from unity, $\sigma$, of the
overlap squared~of exact and variational eigenstates, the (appropriately
normalized) expectation values $\nu$ of the virial operator $C$, and the
(normalized) maximum local difference $\omega$ of the momentum-space
representations of exact and variational eigenstates. The physical parameters
are fixed to~the values $m=2\;\mbox{GeV}$ for the particle mass and
$a=2\;\mbox{GeV}^3$ for the harmonic-oscillator coupling. A simple entry
``0'' indicates that the numerical value is closer to 0 than the rounding
error.}\label{Tab:QVS}\footnotesize
\begin{center}\begin{tabular}{lrlllll}\hline\hline&&\\[-1.5ex]
Quantity&\multicolumn{1}{c}{$d$}&\multicolumn{5}{c}{State}\\[1ex]
\cline{3-7}\\[-1.5ex]&&\multicolumn{1}{c}{1S}&\multicolumn{1}{c}{2S}&
\multicolumn{1}{c}{3S}&\multicolumn{1}{c}{1P}&\multicolumn{1}{c}{1D}\\[1ex]
$n_{\rm r}$&&\multicolumn{1}{c}{0}&\multicolumn{1}{c}{1}&
\multicolumn{1}{c}{2}&\multicolumn{1}{c}{0}&\multicolumn{1}{c}{0}\\
$\ell$&&\multicolumn{1}{c}{0}&\multicolumn{1}{c}{0}&\multicolumn{1}{c}{0}&
\multicolumn{1}{c}{1}&\multicolumn{1}{c}{2}\\[1ex]\hline&&\\[-1.5ex]
$E_{\rm NR}$ [GeV]&&4.12132&6.94975&9.77817&5.53553&6.94975\\[1ex]
$E(m=0)$ [GeV]&&2.94583&5.15049&6.95547&4.23492&5.35237\\[1ex]
\hline&&\\[-1.5ex]
$E$ [GeV]&&3.82493&5.79102&7.48208&4.90145&5.89675\\[1ex]\hline&&\\[-1.5ex]
$\widehat E$ [GeV]&1&4.21624&---&---&6.50936&9.77866\\
&2&3.92759&8.10850&---&5.24154&7.18242\\
&3&3.92684&6.40425&14.4358&4.98863&6.32228\\
&10&3.82530&5.82005&7.64092&4.90409&5.90122\\
&25&3.82494&5.79114&7.48290&4.90149&5.89681\\[1ex]\hline&&\\[-1.5ex]
$\varepsilon$ [cf. Eq.~(\ref{eq:REE})]&1&0.1023&---&---&0.3280&0.6583\\
&2&0.0268&0.4002&---&0.0694&0.2180\\
&3&0.0266&0.1059&0.9294&0.0178&0.0722\\
&10&0.0001&0.0050&0.0212&0.0005&0.0008\\
&25&0&0&0.0001&0&0\\[1ex]\hline&&\\[-1.5ex]
$\sigma$ [cf. Eq.~(\ref{eq:1-Oý})]&1&0.09618&---&---&0.36144&0.65587\\
&2&0.02375&0.43693&---&0.09001&0.34398\\
&3&0.02280&0.13878&0.83034&0.01918&0.12705\\
&10&0.00003&0.00560&0.03727&0.00040&0.00061\\
&25&0&0&0.00008&0&0\\[1ex]\hline&&\\[-1.5ex]
$\nu$ [cf. Eq.~(\ref{eq:REV})]&1&$-0.6120$&---&---&$-0.8328$&$-0.9074$\\
&2&$+0.0308$&$-0.8666$&---&$-0.5103$&$-0.7483$\\
&3&$-0.0187$&$-0.4577$&$-0.9525$&$-0.1275$&$-0.5197$\\
&10&$-0.0034$&$-0.0167$&$-0.1895$&$-0.0016$&$-0.0175$\\
&25&0&$-0.0001$&$+0.0001$&0&0\\[1ex]\hline&&\\[-1.5ex]
$\omega$ [cf. Eq.~(\ref{eq:REW})]&1&$+0.9277$&---&---&$+0.7541$&$+$1.0578\\
&2&$-0.00754$&$+2.4577$&---&$+0.3598$&$+0.7262$\\
&3&$-0.01049$&$+0.2265$&$+4.3252$&$+0.1146$&$+0.4209$\\
&10&$-0.00867$&$+0.0587$&$+0.2592$&$+0.0183$&$+0.0186$\\
&25&$+0.00003$&$-0.0017$&$+0.0002$&$+0.0004$&$+0.0003$\\[1ex]
\hline\hline\end{tabular}\end{center}\end{table}\normalsize\clearpage
\item The deviation from unity, $\sigma$, of the modulus squared of the
overlap $S$ of exact~and variational eigenstates defined in
Eq.~(\ref{eq:OTE}),\begin{equation}\sigma\equiv1-|S|^2\label{eq:1-Oý}\
,\end{equation}is clearly confined to the range$$0\le\sigma\le1\ ;$$the lower
bound applies when the variational eigenstate $|\varphi\rangle$ {\em
agrees\/} with the~exact eigenstate $|\chi\rangle$, while the upper bound
applies when the variational eigenstate~$|\varphi\rangle$ is {\em
orthogonal\/} to the exact eigenstate $|\chi\rangle$. Just for comparison,
the Eckart bounds (\ref{eq:EC}) on $\sigma$, unfortunately available only for
the ground state, are given in Table~\ref{Tab:EB}. In our case, these bounds
turn out to be always larger than the actual numerical values of the
ground-state overlap by at least a factor 2.\small\begin{table}[ht]\caption[]
{\small Eckart bounds on the difference to unity of the overlap squared
$\sigma$ of Eq.~(\ref{eq:1-Oý})~(which apply only to the ground state,
identified by vanishing radial and orbital-angular-momentum quantum numbers
$n_{\rm r}=\ell=0$) for increasing dimension $d=1,2,3,10,25$ of the trial
subspace}\label{Tab:EB}\footnotesize\begin{center}\begin{tabular}{rl}
\hline\hline&\\[-1.5ex]\multicolumn{1}{c}{$d$}&$\displaystyle\frac{\widehat
E_0-E_0}{E_1-E_0}$\\[2.5ex]\hline&\\[-1.5ex]
1&0.19903\\2&0.05222\\3&0.05183\\10&0.00019\\25&0.00001\\[1ex]
\hline\hline\end{tabular}\end{center}\end{table}\normalsize\item The use of
the expectation values of the commutators $[G,H]$ with respect to~the
variational eigenstates $|\varphi\rangle$ is illustrated for the particular
example of the dilation generator $G$ defined in Eq.~(\ref{eq:DG}), by
considering (suitably normalized) expectation values
$\langle\varphi|C|\varphi\rangle$ of the virial operator $C$ given in
Eq.~(\ref{eq:VO}):
\begin{equation}\nu\equiv\frac{\langle\varphi|C|\varphi\rangle}
{\langle\varphi|{\bf x}\cdot\frac{\partial}{\partial{\bf x}}V({\bf x})
|\varphi\rangle}=\frac{\langle\varphi|{\bf p}\cdot\frac{\partial}
{\partial{\bf p}}T({\bf p})|\varphi\rangle}{\langle\varphi|{\bf
x}\cdot\frac{\partial}{\partial{\bf x}}V({\bf x}) |\varphi\rangle}-1\
.\label{eq:REV}\end{equation}\item Finally, the normalized maximum difference
of the normalized momentum-space representations $\tilde\varphi({\bf p})$ and
$\tilde\chi({\bf p})$ of variational eigenstate $|\varphi\rangle$ and exact
eigenstate $|\chi\rangle$, respectively, i.e., the maximum pointwise relative
error in momentum space,\begin{equation}\omega\equiv
\frac{\displaystyle\max_{\bf p}[\tilde\varphi({\bf p})-\tilde\chi({\bf
p})]}{\displaystyle\max_{\bf p}\tilde\chi({\bf
p})}\label{eq:REW}\end{equation} is listed.\end{enumerate}Note that the only
measure for the accuracy of approximate eigenstates $|\varphi\rangle$ which
does not require any information other than the one provided by the
variational technique is $\nu$, the (normalized) expectation values of the
commutator $[G,H]$ with respect to~$|\varphi\rangle$. Inspection of
Table~\ref{Tab:QVS} reveals that $\nu$ represents indeed a sensitive measure
of quality:~for increasing trial-space dimension $d$ it converges to zero at
roughly the same rate as both energy and overlap error, $\varepsilon$ and
$\sigma$, but makes more sense than a pointwise error like~$\omega$.

\section{Summary and Conclusions}This work has been devoted to a systematic
investigation of the quality of solutions~of the eigenvalue problem for some
(semibounded, self-adjoint) operator $H$ by variational methods. Clearly,
even if the overlap of a variationally derived approximate eigenstate and the
corresponding exact eigenstate is close to unity, this does not
necessarily~imply that expectation values of given operators (among which
this operator $H$ which defines the particular eigenvalue problem under
consideration is just one) with respect to these two eigenstates are equal to
the same degree of accuracy. For instance, a given~operator might probe
regions in representation space where the difference between approximate and
exact solutions is of less importance for the overlap. (Only for functions
$f(H)$ of~$H$, the difference between approximate and exact expectation
values of $f$ is, in general, of second order in the deviation of the
approximate eigenstate from the exact eigenstate.) In view of this, various
measures for the accuracy of approximate eigenstates have been discussed.
Apart from the possible exception of some more or less pathological special
cases (cf. our corresponding remarks in Sec.~\ref{Sec:MQ}), the vanishing of
the expectation values of the commutator of $H$ and any other well-defined
operator, taken with respect to~the approximate eigenstates, provides a
useful set of criteria for estimating the significance of the variational
solution. This has been illustrated by considering the commutator~of the
Hamiltonian of the spinless Salpeter equation---which represents the first
step from the nonrelativistic Schr\"odinger formalism towards incorporation
of relativistic effects and which is used for the semirelativistic
description of the spin-averaged mass spectra of bound states of
``constituent'' quarks within the framework of potential models (for analyses
of these attempts see,~e.g., Refs.~\cite{Lucha91:BSQ,Lucha92:,Lucha92})---and
the generator of dilations.

\section*{Acknowledgements}We would like to thank H.~Narnhofer for
stimulating discussions and a critical reading of the manuscript.

\appendix\section{The Relativistic Harmonic-Oscillator Problem}\label{App:RHOP}
In spite of the fact that it is somewhat off the main stream of the present
investigation, let us demonstrate how, with only slightly more effort, the
bounds on the energy levels of the relativistic harmonic oscillator derived
in Sec.~\ref{Sec:SSE} may be considerably improved:\footnote{\ We thank the
referee, presumably Andr\'e Martin, for his ``suggestion'' to include this
discussion.}\begin{itemize}\item By considering the square of the (obviously
self-adjoint) operator $T-\mu$, where~$\mu$ is an arbitrary real parameter
(with the dimension of mass), one obtains a set~of operator inequalities for
the kinetic energy $T$ given in Eq.~(\ref{eq:RKE})
\cite{Martin88,Lucha96:AUB,Lucha98O,Lucha98D,Lucha98R}:$$T\le\frac{{\bf
p}^2+m^2+\mu^2}{2\,\mu}\quad\mbox{for all}\ \mu>0\ .$$These inequalities
immediately translate into upper bounds on the energy levels $E_k,$
$k=0,1,2,\dots,$ of the relativistic harmonic-oscillator problem, which
involve the corresponding nonrelativistic energy eigenvalues $E_{k,{\rm NR}}$
recalled in Eq.~(\ref{Eq:NREL}):$$E_k\le E_{k,{\rm upper}}(\mu)\equiv
\frac{m^2+\mu^2}{2\,\mu}+\sqrt{\frac{m}{\mu}}\left(E_{k,{\rm
NR}}-m\right),\quad\mu>0\ .$$\item On the other hand, by taking into account,
for instance, the obvious positivity~of the square of the (self-adjoint)
operator ${\bf p}^2\,\xi^2-m^2\,(1-\xi^2)$ for an arbitrary~real parameter
$\xi$, it is straightforward to convince oneself of the validity of the set~of
operator inequalities$$T\ge|{\bf p}|\,\sqrt{1-\xi^2}+m\,\xi\ ,\quad 0\le\xi\le
1\ .$$Remembering the energy eigenvalues $E_k(m=0)$ of the Hamiltonian
$H(m=0)$ corresponding to vanishing particle mass $m$ introduced in
Sec.~\ref{Sec:SSE}, these operator inequalities may be reformulated in a
straightforward way as lower bounds on the energy levels $E_k,$
$k=0,1,2,\dots,$ of the relativistic harmonic-oscillator problem:$$E_k\ge
E_{k,{\rm lower}}(\xi)\equiv\xi\,m+(1-\xi^2)^{1/3}\,E_k(m=0)\ ,\quad
0\le\xi\le 1\ .$$\end{itemize}Numerical optimization of these bounds with
respect to the parameters $\mu$ and $\xi$ then yields the improved upper and
lower bounds on the energies of the lowest-lying states of the relativistic
harmonic oscillator listed in Table~\ref{Tab:IULB}.

\small\begin{table}[ht]\caption[]{\small Improved upper and lower bounds on
the lowest energy levels of the semirelativistic Hamiltonian $H=\sqrt{{\bf
p}^2+m^2}+V(r)$ with harmonic-oscillator potential $V(r)=a\,r^2$, for states
with radial quantum number $n_{\rm r}=0,1,2$ and orbital angular momentum
$\ell=0,1,2$~(denoted by 1S, 2S, 3S, 1P, and 1D in usual spectroscopic
notation), for a particle mass of $m=2\;\mbox{GeV}$ and a harmonic-oscillator
coupling strength of $a=2\;\mbox{GeV}^3$.}\label{Tab:IULB}\footnotesize
\begin{center}\begin{tabular}{llllll}\hline\hline&&\\[-1.5ex]
Optimized Bounds&\multicolumn{5}{c}{State}\\[1ex]
\cline{2-6}\\[-1.5ex]&\multicolumn{1}{c}{1S}&\multicolumn{1}{c}{2S}&
\multicolumn{1}{c}{3S}&\multicolumn{1}{c}{1P}&\multicolumn{1}{c}{1D}\\[1ex]
$n_{\rm r}$&\multicolumn{1}{c}{0}&\multicolumn{1}{c}{1}&
\multicolumn{1}{c}{2}&\multicolumn{1}{c}{0}&\multicolumn{1}{c}{0}\\
$\ell$&\multicolumn{1}{c}{0}&\multicolumn{1}{c}{0}&\multicolumn{1}{c}{0}&
\multicolumn{1}{c}{1}&\multicolumn{1}{c}{2}\\[1ex]\hline&&\\[-1.5ex]
$\displaystyle\min_\mu E_{\rm upper}(\mu)$
[GeV]&3.89851&5.99081&7.80385&4.98990&5.99081\\[2ex]
$\displaystyle\max_\xi E_{\rm lower}(\xi)$
[GeV]&3.75534&5.68007&7.36329&4.85577&5.86508\\[2ex]
\hline\hline\end{tabular}\end{center}\end{table}\end{document}